\documentclass[prb,twocolumn,showpacs,floatfix]{revtex4}
\usepackage{graphicx}
\usepackage{amsmath}
\usepackage{amssymb}
\usepackage{braket}
\usepackage[english]{babel}
\usepackage[utf8]{inputenc}
\usepackage{blindtext}
\usepackage{booktabs}
\usepackage{multirow}
\usepackage{subfigure}
\usepackage{cases}
\usepackage{notes2bib}

\newcommand{\head}[1]{\textnormal{\textbf{#1}}}

\newcommand{\AlTwoOThree}{Al$_{\mathrm{2}}$O$_{\mathrm{3}}$}

\newcommand{\CaO}{CaO}

\newcommand{\SiOTwo}{SiO$_{\mathrm{2}}$}
\newcommand{\SiThreeNFour}{Si$_{\mathrm{3}}$N$_{\mathrm{4}}$}

\begin{document}


\title{Electron energy loss spectroscopy of wall charges in plasma-facing dielectrics}


\author{E. Thiessen, F. X. Bronold, and H. Fehske}
\affiliation{Institut für Physik, Universität Greifswald, 17489 Greifswald, Germany}


\date{\today}

\begin{abstract}
We propose a setup enabling electron energy loss spectroscopy to determine
the density of the electrons accumulated by an electro-positive dielectric in 
contact with a plasma. It is based on a two-layer structure inserted into 
a recess of the wall. Consisting of a plasma-facing film made out of the 
dielectric of interest and a substrate layer the structure is designed 
to confine the plasma-induced surplus electrons to the region of the
film. The charge fluctuations they give rise to can then be read out from the
backside of the substrate by near specular electron reflection. To obtain in 
this scattering geometry a strong charge-sensitive reflection maximum due to 
the surplus electrons the film has to be most probably pre-n-doped and sufficiently 
thin with the mechanical stability maintained by the substrate. We demonstrate 
the feasibility of the proposal by calculating the loss spectrum for an
\AlTwoOThree\ film on top of a \CaO\ layer. We find a reflection maximum 
strongly shifting with the density of the surplus electrons and suggest to 
use it for its diagnostics.
\end{abstract} 


\maketitle

\section{Introduction}
The most fundamental manifestation of the interaction of a plasma with a solid is
the formation of an electric double layer consisting, respectively, of an electron-depleted
and electron-rich space charge region on the plasma and the solid side of the 
interface~\cite{BF17}. It arises because electrons are deposited more efficiently onto or
into the surface, depending on its electronic structure, than they are extracted from it by 
neutralization/de-excitation of ions/radicals~\cite{LL05}. Since the beginning of gaseous 
electronics~\cite{LM24} it is of course known that an electric double layer is formed at 
plasma-solid interfaces. Yet a microscopic understanding of the solid-based part of the 
double layer is still missing, mostly because of the limitations of the diagnostics for 
surface charges and because it was so far--perhaps--not essential for the success of 
plasma physics and technology. Continuing progress in the miniaturization of integrated 
microdischarges~\cite{EP13,CH08}, however, driven by the desire to combine solid-state and 
gaseous electronics~\cite{TP17,WT10}, makes the embracing solid structure an integral part 
of the plasma-device. In these structures the solid- and plasma-based charge dynamics are 
intimately linked. A complete understanding of the discharge requires thus to upgrade plasma 
diagnostics by techniques which provide also a view on the charge dynamics inside the 
plasma-facing solid. 

There exist a number of techniques to estimate the charge accumulated by plasma-facing
solids. Electric probes~\cite{KA80}, surface potential measurements~\cite{LL08,OS08,RB03}, 
opto-mechanical devices based on the reflection of a laser by a cantilever~\cite{PF96}, and 
the Pockels effect of an electro-optic crystal~\cite{VSO18,TB14,BW12,GC08,SY07,JB05} have been 
employed for that purpose. However, with the exception of the Pockels effect measurements, the 
methods are rather invasive. In addition, they are limited to measuring the total charge 
accumulated by the plasma-facing solid. How the charge is distributed normal to the 
plasma-wall interface cannot be determined. Information about the charge dynamics inside 
the solid can also not be obtained by these methods. 

To overcome the limitations of the existing methods we recently proposed 
infrared attenuated reflection (IR-ATR) spectroscopy as a tool for gaining access to the 
surplus charges in a dielectric exposed to a plasma~\cite{RBF18}. The 
proposal relies on a layered structure supporting a Berreman mode in the infrared which 
turns out to be rather charge-sensitive. Combined with a self-consistent description of the 
electric double layer at the plasma-solid interface the method has the potential to provide 
not only the total charge deposited into the solid (which we demonstrated by an
exploratory calculation~\cite{RBF18}) but also its spatial distribution inside the solid.
\begin{figure}[b]
        \centering
        \includegraphics[width=0.9\linewidth]{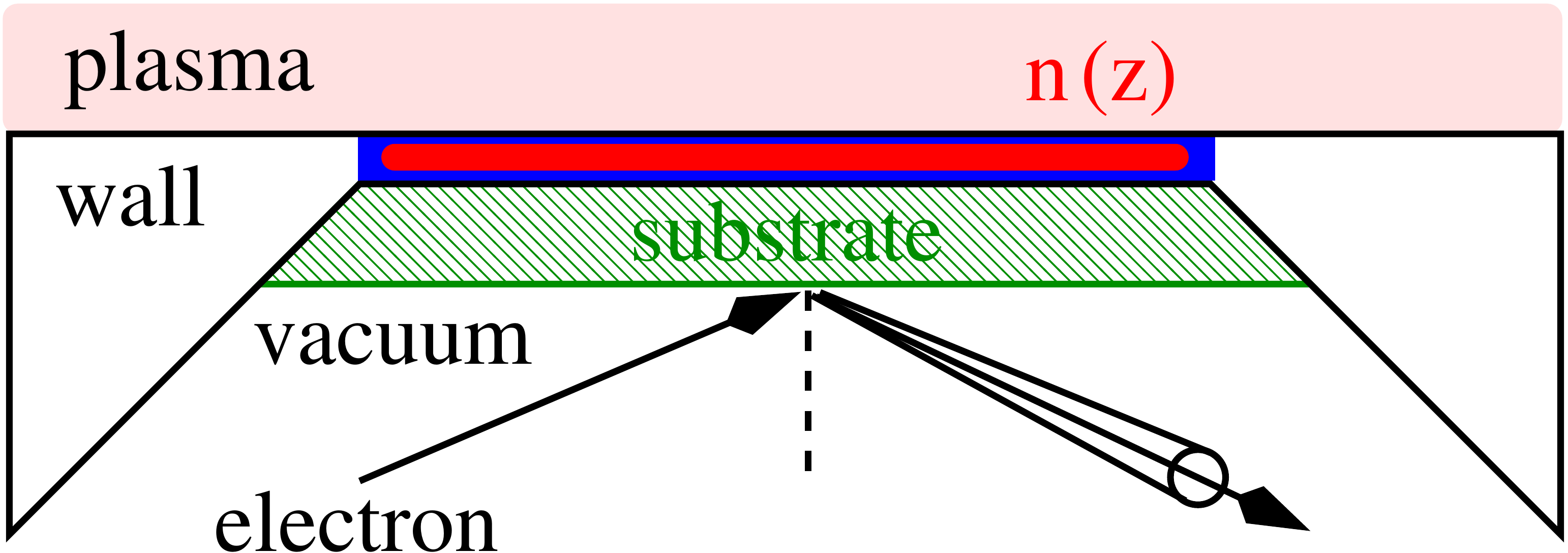}
        \caption{Illustration of the proposed setup for measuring the plasma-induced wall charge
         by electron energy loss spectroscopy. The idea is to confine the wall
         charge $n(z)$ in a thin film, stabilize it by a substrate, and read out the charge
         information with an electron beam applied from the backside.
        }
        \label{testbed}
\end{figure}

Another experimental technique of solid state physics--electron energy loss spectroscopy
(EELS)~(see Refs.~\cite{IM82,LVL85,TL87,I94,Rizzi97,Lueth15} for a general 
introduction)--could be perhaps also used as a diagnostics for the electrons
accumulated by a plasma-facing solid. EELS is an electron reflection technique 
where the probing electron couples to the dipole fields of the charge fluctuations inside 
the solid. As a result it looses energy as well as momentum and scatters a bit off the 
specular direction. The cross section for the near specular reflection contains thus 
information about charge fluctuations inside the solid. 

In a number of experiments it was shown that EELS can be used to determine parameters
characterizing the inhomogeneous electron gases formed at semiconductor 
surfaces~\cite{L88,L86,RL84,L83,MV04}. If it were not for the fact that the plasma prevents 
applying the probing electron beam directly to the plasma-solid interface it would be clear 
that EELS can be used as a wall charge diagnostics in plasma applications. To enable EELS 
to measure the charge accumulated by a plasma-facing solid an indirect setup has to be used. 
It is the purpose of this paper to describe such a setup and to demonstrate by a model 
calculation its feasibility as a testbed for investigating the charging of solids
exposed to a plasma. 

As a first step we focus on a setup measuring the total charge accumulated by an electro-positive 
dielectric in contact with a plasma, leaving modifications required for an analysis of 
the depth profile for future work. The setup is shown in Fig.~\ref{testbed}. It consists 
of a film made out of the dielectric of interest in contact with the plasma and supported 
by a substrate layer. The electron beam supposed to read out the charge information 
is applied from the side opposite to the plasma-solid interface. Interference with 
the plasma is thus excluded on the expense of using a thin film structure. For the 
setup to work it has to be designed in a particular manner. Our calculations 
indicate the probing electron beam to be sensitive to the fluctuations of the 
plasma-induced surplus electrons in the film if the thickness of the whole structure 
is in the sub-100 nm range. In addition we found it advantageous to confine the surplus 
electrons coming from the plasma to the region of the film by the line-up of the 
conduction band edges of the film and the substrate. To compensate for the loss 
of signal strength due to the substrate we suggest moreover to pre-n-dope the plasma-facing 
film. The last measure is not of principal importance. It is only required for 
making the signal detectable with EELS instrumentation currently available. 

\begin{figure}[t]
        \centering
        \includegraphics[width=0.9\linewidth]{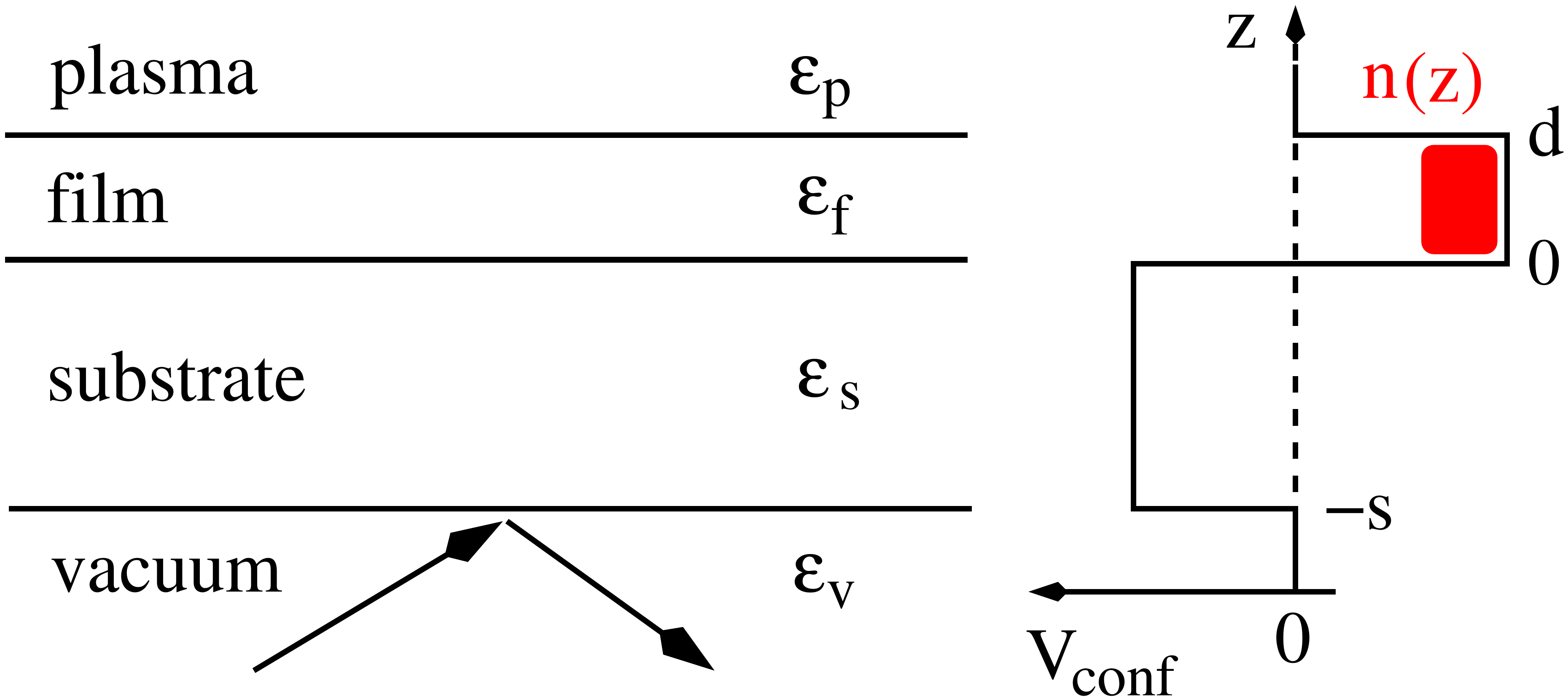}
        \caption{Geometry of the structure we investigate. It consists of an
         electro-positive film with thickness $d$ and an electro-negative
         substrate with thickness $s$ embedded between a plasma and a vacuum.
         The layers as well as the plasma and the vacuum are characterized by
         background dielectric functions $\varepsilon_i$ with $i=p,f,s,v$ as 
         indicated and the electron beam is applied from the vacuum side. On the right 
         is plotted the potential profile confining electrons to the region of the film. 
         It arises from the positive and negative electron affinities $\chi$ of the film 
         and the substrate, respectively, and is essential for the operation of the charge 
         measuring device we propose.
        }
        \label{geometry}
\end{figure}

The outline of the rest of the paper is as follows. In the next section we calculate the 
cross section for near specular electron reflection from a two-layer structure in contact 
with a plasma. Due to the confinement of the charge to a thin film the de Broglie 
wavelength of the electrons is on the same order as the screening length forcing us 
to employ the nonlocal response theory of Mills and coworkers~\cite{Mills75,EM87,SM89}.
Numerical results for an \AlTwoOThree\ film on top of a \CaO\ layer, which is a material 
combination meeting the requirements listed above, are then presented in Section \ref{Results}.
For film thicknesses on the order of a few 10s nanometers and background dopings on 
the order of $10^{18}\,\mathrm{cm^{-3}}$ we find a maximum in the reflection cross 
section due to the collective excitation of the total number of electrons in the film 
which is sufficiently strong to be detectable and charge-sensitive to serve as 
a diagnostics for the density of the film's surplus electrons coming from the plasma. 
Section IV concludes the presentation by summarizing its main points.

\section{Theoretical background}

We consider EELS in an unusual geometry where the probing electron 
approaches a layered structure of finite width from the side opposite
to the interface of interest, which--in our case--is the interface 
between a plasma and an upper most layer of a stack of materials. As illustrated 
in Fig.~\ref{geometry}, the structure and its embedding are characterized by a set 
of dielectric functions $\varepsilon_i$ with $i=v,s,f,p$ and a potential profile 
$V_{\rm conf}(z)$ accounting for the offsets of the layer's conduction band 
edges from the potential just outside the structure (electron affinities $\chi$). 
The energy loss we want to detect and use as a charge diagnostics arises from the 
density $n(z)$ of electrons accumulated in the film next to the plasma. For 
simplicity we characterize the plasma by a dielectric function $\varepsilon_p$ 
which we moreover set at the end to $\varepsilon_v$ neglecting thereby 
charge fluctuations inside the plasma. To take them into account is beyond the 
scope of the present work. It constitutes however no principal problem. 

\subsection{Cross section for EELS} 
Using the coordinate system of Fig.~\ref{geometry} and assuming charge fluctuations
to arise solely from the stack of materials the microscopic approach developed by 
Mills~\cite{Mills75} gives for the momentum-integrated EELS cross section 
\begin{align}
\frac{\mathrm{d}S}{\mathrm{d}\omega} = \frac{2e^2}{\pi \hbar} |R_I|^2 
\int_D \mathrm{d}^2 q_\parallel R(\vec{q}_\parallel, \omega) P(\vec{q}_\parallel, \omega) \; 
\label{scatt_eff}
\end{align}
with $|R_I|^2$ the probability for quantum-mechanical reflection from the vacuum-substrate 
interface, which we set in the following to unity, the surface-rider function 
\begin{align}
R(\vec{q}_\parallel, \omega) = \frac{v_\perp^2}
{[ v_\perp^2 q_\parallel^2 + (\omega - \vec{v}_\parallel \cdot \vec{q}_\parallel)^2]^2}\; ,
 \label{rider_term}
\end{align}
accounting for the scattering kinematics, where $\vec{v}_\parallel$ and $\vec{v}_\perp$ are 
the electron's velocity components parallel and perpendicular to the surface, and the 
loss function ($a=|\vec{a}|$ for any vector $\vec{a}$)
\begin{widetext}
\begin{align}
P(\vec{q}_\parallel, \omega) = \frac{e^2}{\hbar} \int \mathrm{d}^2x_\parallel  
\int_{-\infty}^{\infty} \mathrm{d}t \; e^{\mathrm{i} 
\vec{q}_\parallel \cdot \vec{x}_\parallel - \mathrm{i} \omega t} 
\int_{-s}^{d} \mathrm{d}z \int_{-s}^{d}  \mathrm{d}z' e^{-q_\parallel(z+z'+2s)} \braket 
{\delta \rho^\dagger(\vec{x}_\parallel, z';t) \delta\rho(0, z;0)}_T \; ,
\label{LossFctGen}
\end{align}
\end{widetext}
taking the transfer of energy $\hbar\omega$ and lateral momentum $\vec{q}_\parallel$ parallel 
to the $xy$-plane from the electron to the solid structure into account. It forces the out-going 
branch of the scattering trajectory to deviate a bit from the specular direction and carries 
therefore the information we are looking for.

To the EELS signal not all lateral momenta $\vec{q}_\parallel$ contribute to the cross section. 
Switching to cylindrical coordinates with the polar angle $\phi$ measured with respect to 
$\vec{v}_\parallel$ one realizes that the integration domain $D$ relevant for EELS is bounded 
by the ellipse equation~\cite{LVL85},
\begin{align}
\frac{1}{[q_\parallel^c(\phi)]^2} = \frac{\sin^2(\phi)}{[q_\parallel^{\mathrm{max}}]^2}
+ \frac{\cos^2(\phi)}{[q_\parallel^{\mathrm{max}}\cos(\phi_i)]^2}\; ,
\label{ellipse}
\end{align}
where $\phi_i$ is the incident angle of the probing beam with respect to the surface 
normal, $q_\parallel^{\rm max}=q_{\rm dB}\phi_a$, with $q_{\rm dB}=\sqrt{2E_0m}/\hbar$ 
the de Broglie wave number of the scattering electron incoming with energy $E_0$ and 
$\phi_a$ half of the acceptance angle of the detector. 

The essential part of the cross section is the loss function $P(\vec{q}_\parallel,\omega)$
which via~\eqref{LossFctGen} is related to the charge-charge correlation function
$\braket{\delta \rho^\dagger(\vec{x}_\parallel, z',t) \delta\rho(0, z,0)}_T$, where
the operator $\delta \rho(\vec{x}_\parallel, z',t)$ describes the charge fluctuations
arising in the spatial region for which $-s<z<d$ and the brackets denote the thermodynamical
average taken over this domain at temperature $T$. Clearly this is an approximation. It
assumes that the plasma-facing structure is in thermal equilibrium. In reality
this is not the case. But at this early stage of exploring EELS as a charge
diagnostics the equilibrium assumption seems justified. Including the non-equilibrium      
aspects would unnecessarily mask the basic idea we want to convey and will be subject 
of future work.

From the integrand in \eqref{LossFctGen} arises a problem for the from-the-back
detection of the EELS signal. Because of the substrate layer the absolute value of the 
argument of the exponential function is even in the most favorite situation 
$q_\parallel s$ with $s$ the thickness of the substrate and~\cite{RL84}
\begin{align}
q_\parallel \simeq \sqrt{\frac{2m_e}{\hbar^2}}
               \frac{\hbar\omega\sin\phi_i}{2\sqrt{E_0}}
\label{q_parallel}
\end{align}
the momentum transferred from the electron to the solid; $\hbar\omega$ is the energy
of the charge fluctuation to be detected and $m_e$ is the electron mass. Compared to a 
system without substrate the EELS signal is hence suppressed by a factor $\exp(-q_\parallel s)$ 
implying for $s\simeq q_\parallel^{-1}$ the intensity to be down by roughly $37\%$. Thin 
substrates are thus required for strong signals but they have to be also mechanically stable 
limiting in practice how thin they can be. 

\subsection{Charge fluctuations}
We are interested in the EELS fingerprint of the electrons residing in the plasma-facing
film. To identify their contribution to~\eqref{scatt_eff} we have to
isolate the film's electronic charge fluctuation $\delta\rho_e$ from the total charge 
fluctuation $\delta\rho$ of the two-layer structure. An elegant scheme to accomplish this, 
due originally to Ehlers and Mills~\cite{EM87}, who used it to describe EELS from space
charge regions at semiconductor surfaces, is based on a consideration of the potential 
fluctuations outside the solid probed by the EELS electron; subsequently Streight and 
Mills~\cite{SM89} applied it also to EELS from semiconducting films. Adopted to 
our situation the potential fluctuations in the region $z<-s$ arising from the charge 
fluctuations in the region $z>-s$ have to be calculated. On the one hand, this can be 
done by using the general expression for the electric potential and expanding the factor
$|\vec{x}-\vec{x}^\prime|^{-1}$ in terms of surface waves~\cite{IM82}. It can be 
however also obtained by solving for a point charge located in the 
film the Poisson equation with boundary conditions appropriate for the two-layer 
structure under consideration. Weighting the result with the distribution 
$\delta\rho_e(\vec{q}_\parallel,z, t)$ of surplus electrons in the film gives then 
an alternative expression for the potential fluctuations which via comparison with the 
former allows to relate $\delta\rho_e(\vec{q}_\parallel, z, t)$ to 
$\delta\rho(\vec{q}_\parallel, z, t)$. 

The first approach, based on the general expression for the electric 
potential in front of the stack, leads to 
\begin{align}
\delta \Phi(\vec{x}, t) = \int \frac{\mathrm{d}^2 q_\parallel}{(2\pi)^2} 
 e^{i\vec{q}_\parallel \cdot \vec{x}_\parallel}  \delta \Phi(\vec{q_\parallel}, z, t)\; 
\label{Pot}
\end{align}
with $z<-s$ and  
\begin{align}
\delta \Phi(\vec{q_\parallel}, z, t) = \frac{2\pi e}{q_\parallel} \int_{-s}^{d} \mathrm{d}z' 
e^{q_\parallel (z-z')} \delta\rho(\vec{q}_\parallel, z', t)\; .
\label{PotKernelGen}
\end{align}

The imbedding strategy, on the other hand, working with the Poisson equation, 
\begin{align}
\nabla \cdot [\varepsilon(z) \nabla  \Phi(\vec{x})]= -4 \pi   e \delta(x') \delta(y') \delta(z'-z),
\label{Poisson}
\end{align}
where the point charge is located at $x'=y'=0$ and $d>z'>0$, that is, inside the film, yields upon 
utilizing the homogeneity in the $xy$-plane, 
\begin{align}
\Phi(\vec{x})=\int \frac{\mathrm{d}^2 q_\parallel}{(2\pi)^2} e^{i\vec{q}_\parallel \cdot \vec{x}_\parallel}
\Phi(q_\parallel, z)
\end{align}
with 
\begin{align}
\Phi(q_\parallel, z) = Ae^{q_\parallel z} + Be^{-q_\parallel z}
\end{align}
and expansion coefficients $A$ and $B$ determined from the boundary conditions appropriate 
for the $z$-dependent dielectric function
\begin{align}
\varepsilon(z)=
\begin{cases}
\varepsilon_v \;\; &\text{  for  } \;\; z<-s\\
\varepsilon_s   &\text{  for  } \;\; -s<z<0\\
\varepsilon_f  &\text{  for  } \;\; 0<z<d\\
\varepsilon_p   &\text{  for  } \;\; z>d
\end{cases}\; .
\end{align}
For the imbedding strategy to be applicable the imaginary parts of the dielectric functions 
$\varepsilon_i$ have to be of course negligible in the frequency range of interest. Enforcing 
the boundary conditions
\begin{align}
\Phi(q_\parallel, z^-) = \Phi&(q_\parallel, z^+)\\
\varepsilon(z^-) \frac{\mathrm{d}\Phi(q_\parallel, z)}{\mathrm{d}z}\bigg|_{z^-}-& \varepsilon(z^+) 
\frac{\mathrm{d}\Phi(q_\parallel, z)}{\mathrm{d}z}\bigg|_{z^+} \! =
        \begin{cases}
        4\pi e \;\; \mathrm{if} \;\; z=z'\\
        0 \;\;\;\;\;\; \mathrm{else}
        \end{cases}
\end{align}
at $z=-s$, $z=0$, and $z=d$ and setting $B=0$ for $z<-s$ and $A=0$ for $z>d$ we find--after 
weighting the result with the (instantaneous) charge distribution inside the film and Fourier 
transforming the lateral spatial variables--for the potential in the 
region $z<-s$, denoted now again by $\delta\Phi(\vec{q}_\parallel, z, t)$, the expression
\begin{align}
&\delta\Phi(\vec{q}_\parallel, z,t) = 
\cfrac{(2\pi e/q_\parallel) 4 \varepsilon_s}{(\varepsilon_f + \varepsilon_s)(\varepsilon_s+\varepsilon_v)
h(q_\parallel; \varepsilon_s, s; \varepsilon_f, d; \varepsilon_p, \varepsilon_v)} \nonumber \\
&\int_{-s}^{d}\mathrm{d}z' e^{q_\parallel (z-z')}
F(q_\parallel, z'; \varepsilon_f, d; \varepsilon_p)  \delta\rho_e(\vec{q}_\parallel, z',t) \theta(z') \; ,
\label{PotFilm}
\end{align}
with the auxiliary functions 
\begin{align}
h(q_\parallel; \varepsilon_s, s;& \varepsilon_f,d; \varepsilon_p, \varepsilon_v)=
    1+L_{sv}L_{fs}e^{-2 q_\parallel s}\nonumber \\& -  L_{fp}L_{fs}e^{-2 q_\parallel d}
  -  L_{fp} L_{sv}e^{-2 q_\parallel (d+s)}
\end{align}
and
\begin{align}
F(q_\parallel, z; \varepsilon_f, d; \varepsilon_p)=1 + L_{fp} e^{-2 q_\parallel (d-z)}\; ,
\end{align}
where
\begin{align}
L_{ij} = \frac{\varepsilon_i - \varepsilon_j}{\varepsilon_i + \varepsilon_j} \;\; \text{with }\;\; i,j=v,s,f,p \; .
\end{align}

Comparison of~\eqref{PotKernelGen} and \eqref{PotFilm} yields then a relation between 
$\delta\rho(\vec{q}_\parallel, z',t)$ and $\delta\rho_e(\vec{q}_\parallel, z',t)$  
which inserted into~\eqref{LossFctGen} leads after some algebra to the loss function 
\begin{align}
P&(\vec{q}_\parallel, \omega) = 2 e^2 (1+n(\omega)) \nonumber
\\& \left[\frac{4 \varepsilon_s}{(\varepsilon_f + \varepsilon_s)(\varepsilon_s+\varepsilon_v) 
h(q_\parallel; \varepsilon_s, s; \varepsilon_f,d;\varepsilon_p)}\right]^2 \nonumber \\
&\int_{0}^{d} \mathrm{d}z \int_{0}^{d} 
\mathrm{d}z' e^{-q_\parallel(z+z' + 2s)}F(q_\parallel, z; \varepsilon_f, d; \varepsilon_p)  \nonumber \\
&F(q_\parallel, z'; \varepsilon_f, d; \varepsilon_p) \mathrm{Im}\chi(\vec{q}_\parallel, \omega; z, z') \; 
\label{LossFctFilm}
\end{align}
with $n(\omega)$ the Bose distribution function with $\beta=1/(k_B T)$ and
\begin{align}
\chi(\vec{q}_\parallel, \omega; z, z') = \frac{i \theta(\omega)}{\hbar} 
\braket{[ \delta \rho^\dagger_e(\vec{q}_\parallel,z, \omega), \delta \rho_e(\vec{q}_\parallel,z',0)]}_T \;  
\end{align}
the (commutator) density-density response function for the surplus electrons~\cite{EM87,SM89}. 
The calculation of the loss function $P(\vec{q}_\parallel, \omega)$ has thus been reduced to the 
determination of the function $\chi(\vec{q}_\parallel, \omega; z, z')$. That the electrons are 
embedded in a dielectric structure is taken into account in \eqref{LossFctFilm} by the 
functions in front of $\mathrm{Im}\chi(\vec{q}_\parallel, \omega; z, z')$.

\subsection{Density-density response function}
The simplest scheme for obtaining the density-density response function 
$\chi(\vec{q}_\parallel, \omega; z, z')$ relies on the random-phase approximation. 
As in the work of Mills and coworkers~\cite{EM87,SM89} it is based on the integral 
equation,
\begin{align}
\chi(\vec{q}_\parallel, \omega; &z, z') =  \chi_0(\vec{q}_\parallel, \omega; z, z')
\nonumber \\ &-  \int_{0}^{d} \mathrm{d}z''
  K(\vec{q}_\parallel, \omega; z, z'') \chi(\vec{q}_\parallel, \omega; z'', z') \; ,
  \label{int_eq}
\end{align}
where the kernel
\begin{align}
K(\vec{q}_\parallel, \omega; z, z') = \int_{0}^{d} \mathrm{d}z'' \chi_0(\vec{q}_\parallel, \omega; z, z'')
v(q_\parallel; z'', z')
\label{Kernel}
\end{align}
includes the electron-electron interaction $v(q_\parallel; z, z')$ and the irreducible 
particle-hole propagator $\chi_0(\vec{q}_\parallel, \omega; z, z')$. Writing for the interaction 
potential $v(q_\parallel; z, z') = e\Phi(q_\parallel, z, z')$, with $\Phi(q_\parallel, z, z')$
the electric potential at $z$ induced by a point charge at $z'$, where both $z$ and $z^\prime$ 
are inside the film, the solution of the Poisson equation \eqref{Poisson} can be utilized to 
deduce 
\begin{align}
v(q_\parallel; z, z') =  \frac{2\pi e^2}
{q_\parallel \varepsilon_f } \frac{g(q_\parallel; \varepsilon_s,s; \varepsilon_f, d; \varepsilon_p, \varepsilon_v)}
{h(q_\parallel;\varepsilon_s, s; \varepsilon_f,d; \varepsilon_p,\varepsilon_v)}
\label{ww_substr_film}
\end{align}
where 
\begin{align}
        &g(q_\parallel; \varepsilon_s,s; \varepsilon_f, d;\varepsilon_p, \varepsilon_v) =
        e^{-q_\parallel|z-z'|}+ L_{sv}e^{-q_\parallel(z+z' + 2s)} \nonumber \\
        &+L_{fs}e^{-q_\parallel(z+z')}+ L_{fp}e^{-q_\parallel (2 d - (z+z'))}\nonumber \\
        &+ L_{sv}L_{fs}e^{-q_\parallel(2s + |z-z'|)}  +  L_{sv}L_{fp} e^{-q_\parallel(2(s+d) - |z-z'|)}  \nonumber \\&+ L_{fs}L_{fp} e^{-q_\parallel(2d - |z-z'|)} \nonumber \\
        &+L_{sv}L_{fs}L_{fp} e^{-q_\parallel(2(d+s) - (z + z'))} \;.
\end{align}

To complete the construction of the kernel~\eqref{Kernel} we also need the irreducible 
electron-hole propagator $\chi_0(\vec{q}_\parallel, \omega; z, z')$. It is given 
by~\cite{EM87,SM89}
\begin{align}
\chi_0(\vec{q}_\parallel, &\omega; z, z') =  \nonumber \\
&\frac{2}{A} \sum_{\vec{k}_\parallel} \sum_{i,j}\frac{ f(\vec{k}_\parallel, i) - f(\vec{k}_\parallel +  \vec{q}_\parallel, j) }{ \hbar \omega + \mathrm{i}\delta       + E_{\vec{k}_\parallel + \vec{q}_\parallel, j} - E_{\vec{k}_\parallel,i} } \nonumber \\
&\cdot \psi_i^*(z) \psi_i(z')\psi_j(z)\psi_j^*(z') \;,
\label{xi0}
\end{align}
where 
\begin{align}
f(\vec{k}_\parallel, i) = \frac{1}{e^{\beta (E_{\vec{k}_\parallel,i} -\mu )}+1} \; 
\end{align}
is the Fermi distribution function with $\beta=1/(k_B T)$ and $\mu$ is the chemical potential;
$A$ is the quantization area in the $xy$-plane. 

The single electron energies entering~\eqref{xi0} contain the energy of the lateral and the 
vertical motion, 
\begin{align}
E_{\vec{k}_\parallel, i} = \frac{\hbar^2 k_\parallel^2}{2m^*} + \varepsilon_i \; 
\end{align}
with the energy $\varepsilon_i$ belonging to $\psi_i(z)$, the part of the wave 
function describing the perpendicular motion. To obtain $\varepsilon_i$ and $\psi_i(z)$
we assume the film to constitute an infinitely deep potential well. Both quantities 
can then be looked up in textbooks about quantum mechanics~\cite{Gasiorowicz74}. 
It is thus not necessary to list them here. Since the electron affinity of \AlTwoOThree, 
the material we will take for the film, is large and positive, the infinitely deep quantum 
well is a reasonable approximation. Improvements are possible but will be not addressed 
in this paper. The labels $i,j$ are the quantum numbers labeling the 
eigenstates of the well and $\delta$ is a small but finite number preventing numerical 
instability. For the results discussed in the next section $\delta=10^{-5}$. Finally,  
the chemical potential $\mu$ has to be determined. It is related to the surface charge 
density through the condition~\cite{SM88}
\begin{align}
n_s = \frac{m^*}{\pi \hbar^2 \beta} \sum_i \ln(1+e^{-\beta(\varepsilon_i - \mu)})\; .
\label{SurfaceDensity}
\end{align}
Inside the film the electrons are distributed according to 
\begin{align}
n(z) = \frac{m^*}{\pi \hbar^2 \beta} \sum_i \ln(1+e^{-\beta(\varepsilon_i - \mu)})
|\psi_i(z)|^2 \;,
\end{align}
which after integration over $z$ yields~\eqref{SurfaceDensity}, as it should, because 
of the normalization of the wave functions.

\subsection{Remarks concerning the numerics}
The numerical work consists of two major parts: (i) calculating the density-density response 
function $\chi(\vec{q}_\parallel, \omega; z, z')$ by solving the integral equation~\eqref{int_eq}
and (ii) integrating $\chi(\vec{q}_\parallel, \omega; z, z')$ over $z$ and $z^\prime$ as 
specified in \eqref{LossFctFilm} to obtain the loss function $P(\vec{q}_\parallel,\omega)$, which 
is then inserted into~\eqref{scatt_eff} to yield after integrating over $\vec{q}_\parallel$ 
the EELS cross section $dS/d\omega$. 

In order to obtain $\chi(\vec{q}_\parallel, \omega; z, z')$ we first have to construct the 
function $\chi_0(\vec{q}_\parallel, \omega; z, z')$. For that purpose we closely follow Ehlers 
and Mills~\cite{EM87} and adopt their approach to the layered structure we consider. Hence, we 
convert the summation over $\vec{k}_\parallel$ to an integral and rewrite~\eqref{xi0} 
using Green functions. Only one sum over the eigenstate labels $i$ remains then. Due to 
the homogeneity in the $xy$ plane it turns out that all quantities depend only on the absolute 
value of $\vec{q}_\parallel$ which is chosen parallel to the $x$-direction. For the solution of 
the integral equation~\eqref{int_eq} itself we no longer follow Ehlers and Mills~\cite{EM87}. 
Instead we employ the numerical strategy Streight and Mills~\cite{SM89} used in their study
of semiconducting films. They noticed that for a film the numerical work can be greatly 
reduced by integrating out the variable $z'$ which enters~\eqref{int_eq} and~\eqref{LossFctFilm} 
only as a parameter. Instead of solving~\eqref{int_eq} for $\chi(q_\parallel, \omega; z, z')$,
depending as we now known only on $q_\parallel$, we thus solve an integral equation for 
\begin{align}
 X(q_\parallel, \omega;z) = 
  \int_0^d \mathrm{d}z' F (q_\parallel,z'; \varepsilon_f, d;\varepsilon_p) 
   e^{-q_\parallel z'} \chi(q_\parallel, \omega;z,z')~,
 \label{Xfct}
\end{align}
which can be easily derived from~\eqref{int_eq} and efficiently solved by discretization 
and matrix inversion. 


Obtaining $\chi(q_\parallel, \omega; z, z')$ in the manner described is numerically 
the most challenging task. The integrations specified in~\eqref{scatt_eff} 
and~\eqref{LossFctFilm}, on the other hand, can be performed with standard integration 
routines. For the numerical work we used dimensionless variables, measuring energies 
and lengths, respectively, in units of $k_BT$ and $\lambda_*=\sqrt{\hbar^2/2m_e^*k_BT}$, 
where $m_e^*$ is the effective electron mass in the conduction band of the film.

\section{Results}
\label{Results}

\begin{table}[t]
        \begin{center}
                \begin{tabular}{c|c|c}
                        & \CaO & \AlTwoOThree \\\hline
                        $ \varepsilon_\infty$           &  3.3856       & 3.2 \\
                        $ \varepsilon_0$                        &  --           & 9.0   \\
                        $\nu_1[{\rm cm}^{-1}]$      &  300      & 385 \\
                        $f_1$                           &  9            & 0.3 \\
                        $\gamma_1 [{\rm cm}^{-1}]$      &  32           & 5.58 \\
                        $\nu_2[{\rm cm}^{-1}]$      &  --       & 442 \\
                        $f_2$                           &  --       & 2.7 \\
                        $\gamma_2 [{\rm cm}^{-1}]$  &  --           & 4.42 \\
                        $\nu_3[{\rm cm}^{-1}]$      &  --               & 569 \\
                        $f_3$                                   &  --           & 3 \\
                        $\gamma_3 [{\rm cm}^{-1}]$      &  --           & 11.38 \\
                        $\nu_4[{\rm cm}^{-1}]$          &  --           & 635 \\
                        $f_4$                           &  --           & 0.3 \\
                        $\gamma_4 [{\rm cm}^{-1}]$      &  --       & 12.7  
                \end{tabular}
                \caption{Parameters entering~\eqref{eps} for the background dielectric functions
                        of \CaO~\cite{HKS03} and \AlTwoOThree~\cite{P85,B63}, the materials
                        used, respectively, for the substrate and the film of the proposed
                        EELS setup for measuring the wall charge.
                }
                \label{MatPara}
        \end{center}
\end{table}

We now present results for a setup consisting of a plasma-facing \AlTwoOThree\  
film supported by a \CaO\ substrate layer to which an electron beam is applied from 
below with $E_0=5\,\mathrm{eV}$ and $\phi=45^\circ$. The acceptance angle of the detector
is $2\phi_a=2^\circ$.

The combination of materials meets the criteria we impose for the setup to work as a testbed 
for measuring the wall charge and also for our theory to be applicable: (i) charge confinement 
to the film, (ii) dielectric functions with small imaginary parts in the spectral range of 
interest, and (iii) mechanical stability. \AlTwoOThree\ is an electro-positive dielectric with 
electron affinity $\chi = 2.58\,\mathrm{eV}$~\cite{HCC06} whereas \CaO\ is electro-negative 
with $\chi = -0.86\,\mathrm{eV}$~\cite{SS81}. Hence, the surplus electrons 
coming from the plasma will be confined to the film. In the energy range of the
charge fluctuation we probe by EELS, the imaginary parts of the dielectric functions are
moreover very small. The argument enabling us to express the total charge fluctuation 
as a product of a factor describing the background and a factor describing the surplus 
electrons is thus justified. Finally, the mechanical properties of the two materials 
make them suitable for our setup. Their microhardnesses, for instance, are on the same 
order as the ones for \SiOTwo\ and \SiThreeNFour\ which are used as sub-100 nm membranes 
in photoelectron microscopy to withstand pressure gaps at vacuum-liquid interfaces~\cite{TVI16}. 
We expect therefore the \AlTwoOThree/\CaO\ system to allow also the construction of film-substrate 
structures with thicknesses in the sub-100 nm range, as it is required for the 
EELS signal to be detectable in the setup we consider. 

\begin{figure}[t]
        \centering
        \includegraphics[width=1.00\linewidth]{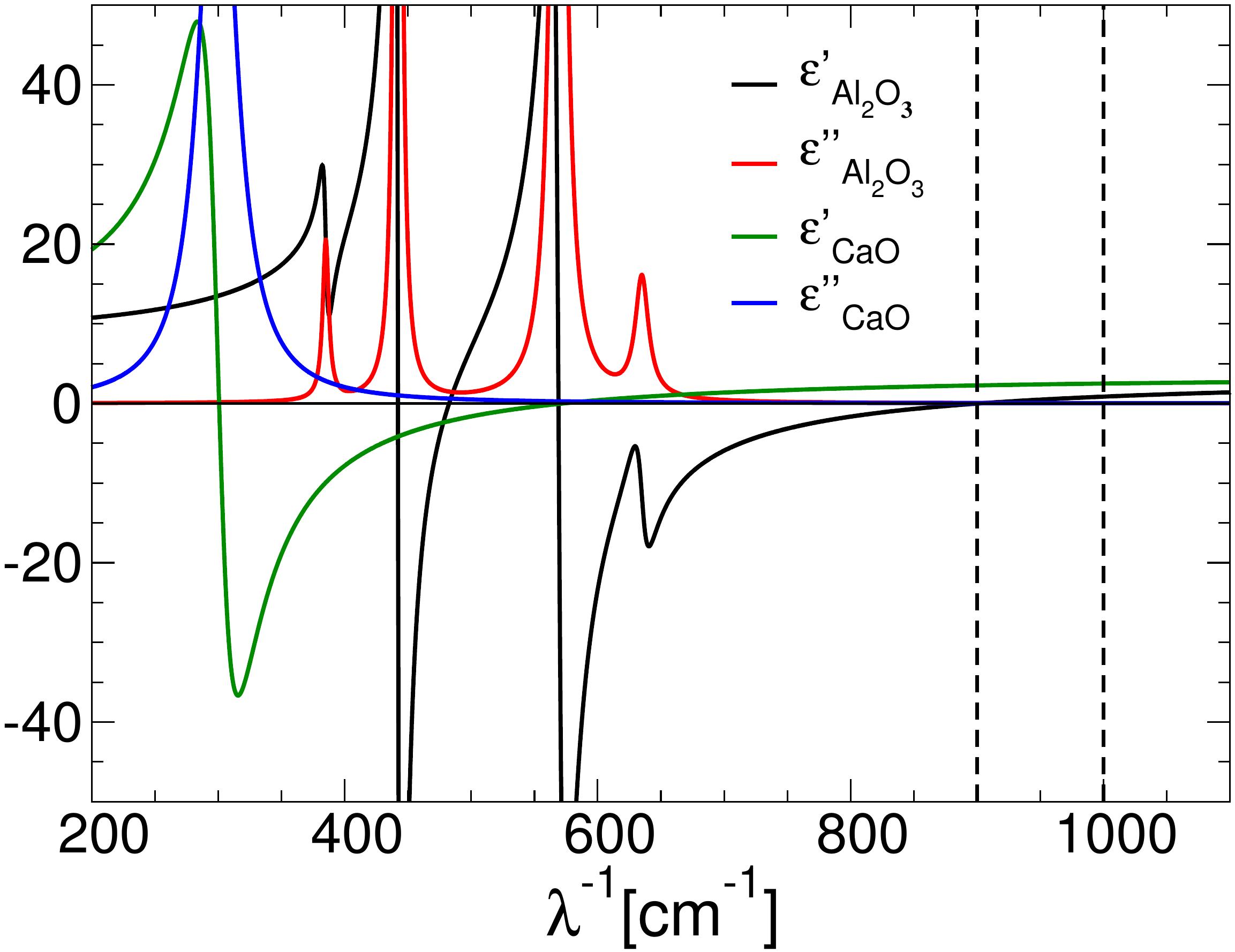}
        \caption{Background dielectric functions for \CaO\ and \AlTwoOThree\ as obtained 
        from~\eqref{eps} using the parameters of Table \ref{MatPara}. The dashed vertical
        lines denote the spectral range of the collective excitation of the electrons inside
        the \AlTwoOThree\ film which we use as a charge diagnostics.
        }
        \label{DielectricFcts}
\end{figure}

The background dielectric functions for \CaO\ and \AlTwoOThree\ are found in the 
literature~\cite{HKS03,P85,B63}. Fitting experimental data to a set of damped 
harmonic oscillators they can be written in the form~\cite{HKS03,P85,B63} 
\begin{align}
\varepsilon(\omega) = \varepsilon_\infty +
\sum_{i=1}^{i=4} \frac{f_i \omega_i^2}{\omega_i^2 - \omega^2 - \mathrm{i}\gamma_i\omega}
\label{eps}
\end{align}
with $\omega_i=2\pi \nu_i=2\pi c/\lambda$. The parameters entering this equation are given 
in Table~\ref{MatPara} for the two materials. Plots of the dielectric functions 
themselves are shown in Fig.~\ref{DielectricFcts}, where we also indicate by vertical 
dashed lines the spectral region where the charge fluctuation occurs whose density 
dependence we utilize for diagnostic purposes. Clearly, in the relevant 
spectral range the dielectric functions are essentially real. In addition
to the dielectric functions $\varepsilon$ and the electron affinities $\chi$ we also 
need the effective electron mass $m_e^*$ in the conduction band of \AlTwoOThree. In 
units of the electron mass $m_e^*=0.4$. The temperature is set to 
$T=300\,\mathrm{K}$ in all calculations and $\varepsilon_p=\varepsilon_v=1$. 

Initially we simulated an undoped stack with thickness $40\,\mathrm{nm}$
(which we consider sufficient for mechanical stability) and indeed found a 
loss peak strongly shifting with 
the density of the surplus electrons accumulated from the plasma and hence 
suitable for our purpose. Unfortunately, the intensity of the peak is
rather small because it arises from a multipole excitation of the electrons.  
In units of the strength of the surface (Fuchs-Kliewer) phonon at the 
vacuum-\CaO\ interface, located at $524\, \mathrm{cm^{-1}}$, 
which we take as a reference strength, the peak height was only around 
$10^{-6}$ for a $10\,\mathrm{nm}$ \AlTwoOThree\ film with a plasma-induced
surface charge density $n_s^p=5\cdot 10^{11}\,\mathrm{cm^{-2}}$ on top of a 
$30\,\mathrm{nm}$ \CaO\ layer. Such a faint signal is most probably 
undetectable by current EELS instrumentation. In a recent application to
nanoplasmonics~\cite{BYT13}, for instance, EELS had a sensitivity of 
$10^{-4}$ in units of the elastic peak. Since the Fuchs-Kliewer phonon 
we use for normalization is typically an order of magnitude weaker than the 
elastic peak, signals should not be weaker than $10^{-3}$ in our units 
to be detectable. 

Although electron counting techniques may advance~\cite{HLL17}, pushing thereby 
the sensitivity limit, we take $10^{-3}$ as a critical value. Measures are thus 
necessary to increase the signal strength up to this value. Increasing
the signal strength by reducing the thickness of the substrate is not
viable because it would threaten the mechanical stability. Another possibility
is to increase the density of the electron gas by pre-n-doping the \AlTwoOThree\ 
film. Due to the pre-doping a loss peak of lower order (and hence more 
intensity) becomes charge sensitive and hence suitable for charge diagnostics. 
It defines also a reference peak, present without surplus electrons from the 
plasma, which should help calibrating the method. The data presented below are 
therefore for a doped \AlTwoOThree\ film. Due to the doping the confinement 
potential is of course no longer simply the potential well arising from the 
electron affinities. The potential is affected by the Coulomb interaction between 
the electrons and should be calculated selfconsistently~\cite{SM88}. But for
demonstrating the basic principle of the charge measurement this is not necessary.
We leave it thus for the future.  

\begin{figure}[t]
        \centering
         \includegraphics[width=1.00\linewidth]{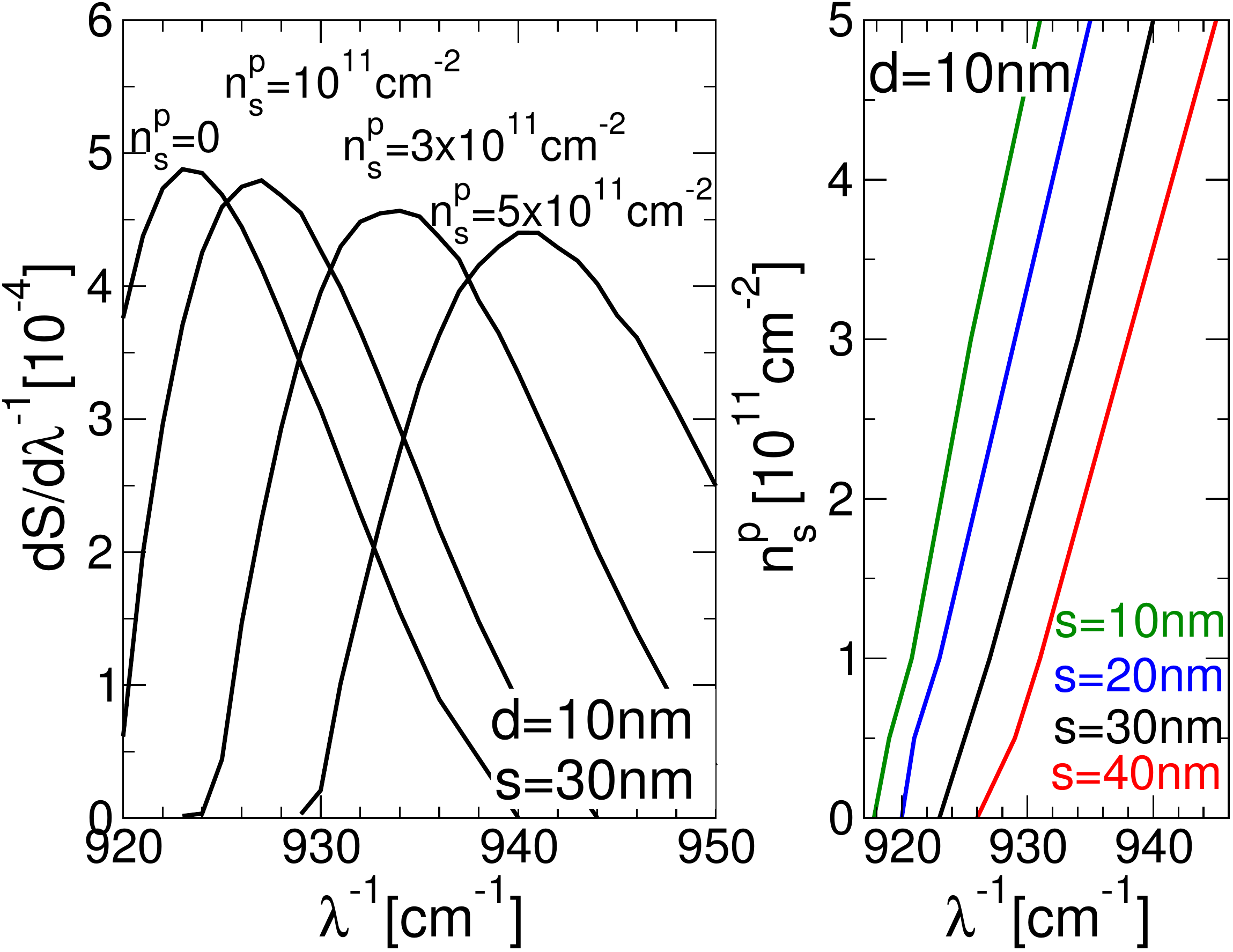}
        \caption{EELS spectra for an \AlTwoOThree/\CaO\ structure at
         $300\, \mathrm{K}$ in the energy range where the resonance is located we 
         use as a diagnostics for the plasma-induced charges in the \AlTwoOThree\ film. 
         The film's thickness and background doping are, respectively, $d=10\,\mathrm{nm}$
         and $n_b^d=10^{18}\, \mathrm{cm^{-3}}$ corresponding to a surface charge density
         $n_s^d=n_b^d d=10^{12}\, \mathrm{cm^{-2}}$. Left panel: Loss peak normalized to the
         strength of the Fuchs-Kliewer phonon of the vacuum-\CaO\ interface as a 
         function of the plasma-induced surface charge density $n_s^p$
         [corresponding to a bulk density $n_b^p=n_s^p/d$] for a substrate with thickness
         $s=30\,\mathrm{nm}$. Right panel: Energetic position of the loss peak as a function
         of $n_s^p$ for four different substrate thicknesses. The film thickness is in 
         all four cases fixed to $d=10\,\mathrm{nm}$.
        }
        \label{FigData1}
\end{figure}

After these remarks let us now turn to the EELS spectrum of a stack with 
a pre-n-doped plasma-facing \AlTwoOThree\ film. Figure~\ref{FigData1} shows 
data for wave numbers where the loss is due to a fluctuation of the electron
gas in a $10\,\mathrm{nm}$ film doped with a bulk electron density 
$n_b^d=10^{18}\, \mathrm{cm^{-3}}$. In the left panel the loss peak is plotted  
as a function of the density $n_s^p$ of additional electrons coming from the 
plasma. Clearly, for a substrate thickness $s=30\,\mathrm{nm}$ the peak has 
not quite the required strength but reducing the substrate thickness a bit 
will push the strength above the critical value as will be discussed in the 
next paragraph. How the peak shifts with $n_s^p$ for different substrate 
thicknesses is plotted in the right panel.  Notice, the position of the peak 
for $n_s^p=0$ depends weakly on $s$ despite the unchanged bulk electron 
density in the film. We attribute this to a small substrate-induced redistribution 
of spectral weight due to changes in the electric field producing slightly different 
loss maxima. For all the chosen values of $s$ the peak shifts nicely with $n_s^p$ and 
is thus well suited for measuring $n_s^p$ by simply recording the spectral position. 
Due to the limited energy resolution of EELS~\cite{Rizzi97}, experimentally detectable 
are only shifts larger than $\Delta \lambda^{-1}=4\,\mathrm{cm^{-1}}$. Hence, charge 
densities $n_s^p > 10^{11}\,\mathrm{cm^{-2}}$ may be measurable by this technique.
But this is also the range expected for plasma-facing dielectrics as can be inferred 
from experimental studies of the charging of dust particles in low-temperature 
plasmas~\cite{KRZ05} and measurements of the wall charge by the Pockels
effect~\cite{SY07,JB05,GC08}.

\begin{figure}[t]
        \centering
        \includegraphics[width=0.9\linewidth]{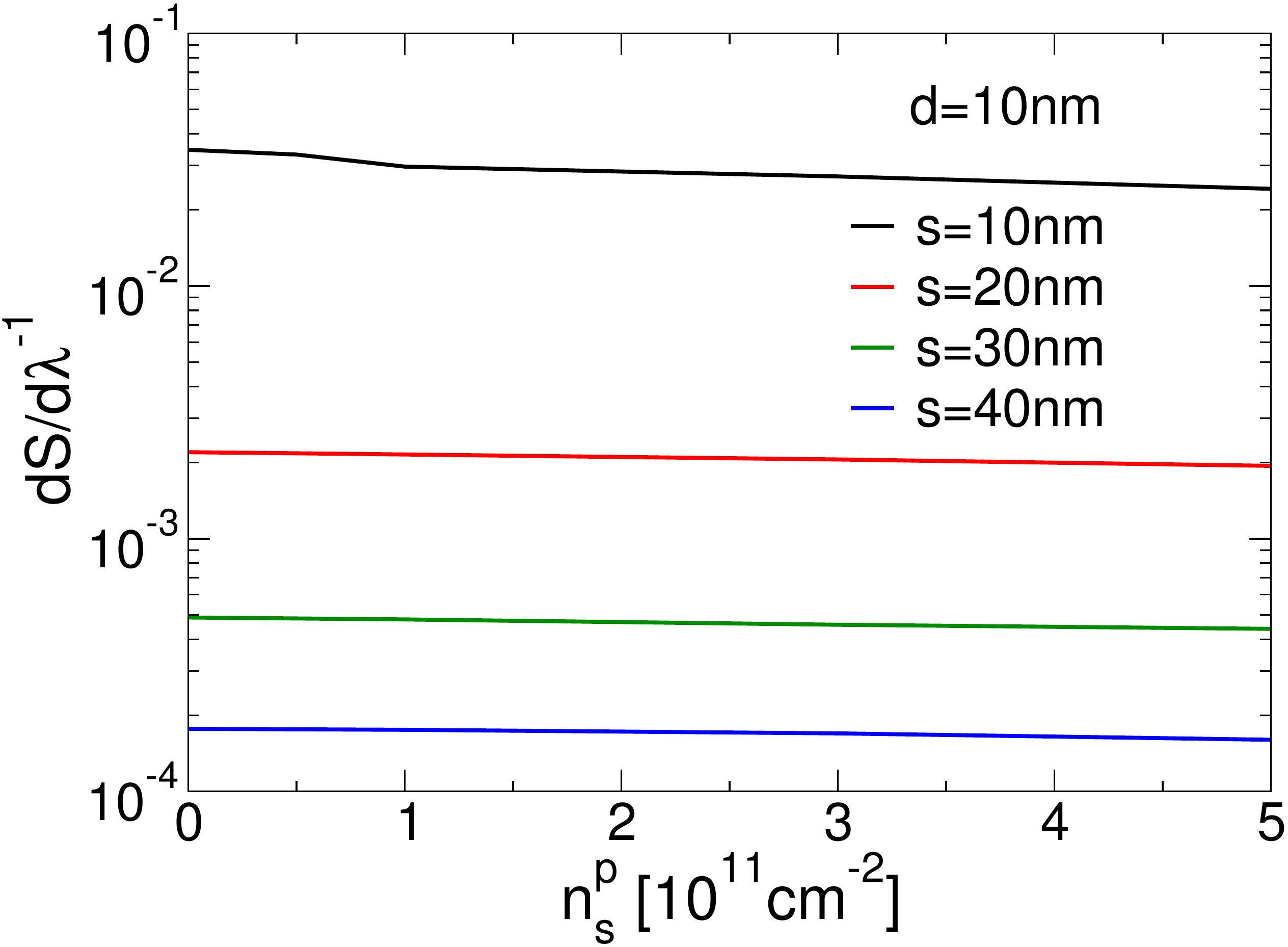}
        \caption{Intensity of the loss peaks shown in Fig.~\ref{FigData1} as a
                 function of $n_s^p$. The thicker the substrate the weaker the
                 loss peak, as expected. The EELS spectrum is normalized to the
                 intensity of the Fuchs-Kliewer phonon at the vacuum-\CaO\ interface.
                 As explained in the main text, we expect the critical strength below 
                 which the signal becomes undetectable to be around 
                 $10^{-3}$ in these units. Hence, the sensitivity of current EELS 
                 instrumentation may be not sufficient for substrates thicker than 
                 $30\,\mathrm{nm}$.  
        }
        \label{FigDataInt}
\end{figure}

The strength of the loss peaks is shown in Fig.~\ref{FigDataInt}. According to the 
considerations presented above we take $10^{-3}$ as the critical strength 
in units of the strength of the Fuchs-Kliewer phonon below which 
the signal cannot be detected anymore. As can be seen, the background doping pushes
the signal strength for $s < 30\,\mathrm{nm}$ above this critical value, leaving the 
system with $s=30\,\mathrm{nm}$ at the margin. Let us at this point however  
caution a bit. In the literature EELS data are mostly given in arbitrary 
units. Our estimate of the critical signal strength is based on one~\cite{BYT13} of 
the few publications where the data are normalized to a particular peak and 
hence estimable from an intensity point of view. It may be possible that electron
detectors used in EELS are in fact more sensitive than we believe. The signal could 
then by accordingly weaker.

\begin{figure}[t]
        \centering
        \includegraphics[width=0.9\linewidth]{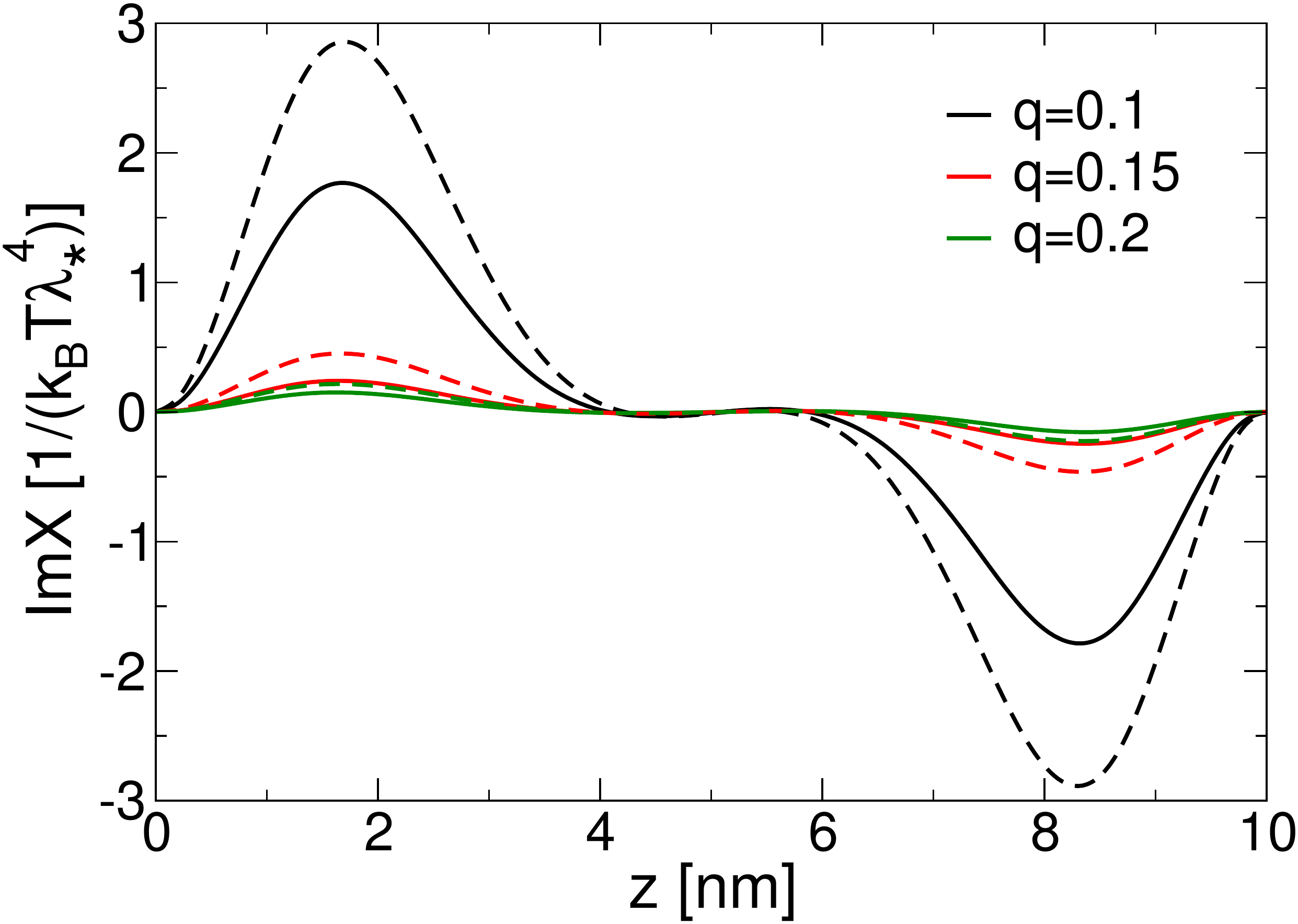}
        \caption{The function $\mathrm{Im} X(q_\parallel,\omega,z)$ for an
         \AlTwoOThree/\CaO\ setup with $d=10\,\mathrm{nm}$, $s=30\,\mathrm{nm}$,
         $n_b^d=10^{18}\,\mathrm{cm^{-3}}$ [corresponding to $n_s^d=10^{12}\,\mathrm{cm^{-2}}$],
         and two values for the density of the plasma-induced surplus electrons: 
         $n_s^p=10^{11}\,\mathrm{cm^{-2}}$ (solid lines) and 
         $n_s^p=5\cdot 10^{11}\,\mathrm{cm^{-2}}$ (dashed lines). The lateral momentum 
         $q_\parallel$ in units of $(\lambda_*)^{-1}=1/\sqrt{\hbar^2/2m_e^*k_BT}$ and the 
         energy $\hbar\omega$ in units of wave numbers making in each case $P(q_\parallel,\omega)$ 
         maximal are given by $(0.1, 927.742\,\mathrm{cm^{-1}})$, $(0.15, 923.538\,\mathrm{cm^{-1}})$, 
         and $(0.2, 921.318\,\mathrm{cm^{-1}})$ for $n_s^p=10^{11}\,\mathrm{cm^{-2}}$ and 
         $(0.1,941.837\,\mathrm{cm^{-1}})$, $(0.15,936.1\,\mathrm{cm^{-1}})$, and 
         $(0.2,932.691\,\mathrm{cm^{-1}})$ for $n_s^p=5\cdot 10^{11}\,\mathrm{cm^{-2}}$.
         For each density the three doubles belong to the domain $D$ over which 
         $P(q_\parallel,\omega)$ has to be integrated according to~\eqref{scatt_eff}. The 
         $z$-dependence of $\mathrm{Im} X(q_\parallel,\omega,z)$ shows that the charge fluctuation 
         giving rise to the peak in $P(q_\parallel,\omega)$ is a surface plasmon 
         localized close to the boundaries of the \AlTwoOThree\ film.
        }
        \label{ChargeFluct}
\end{figure}

In order to understand the physics of the loss peak we are tracking as a function of 
$n_s^p$, we analyzed the spatial structure of the charge fluctuation giving rise to it 
using the procedure developed by Streight and Mills~\cite{SM89}. They noticed that for
$\omega$ residing on the loss peak and $q_\parallel$ fixed to a value contributing to 
the EELS spectrum according to the ellipse equation~\eqref{ellipse} the $z-$dependence 
of the function $\mathrm{Im} X(q_\parallel,\omega,z)$, with 
$X(q_\parallel,\omega,z)$ defined in \eqref{Xfct}, reflects the spatial form of the 
charge fluctuation associated with the peak. Figure~\ref{ChargeFluct} shows this 
function for two different values of $n_s^p$ for a structure with 
$d=10\,\mathrm{nm}$, $s=30\,\mathrm{nm}$, and $n_b^d=10^{18}\,\mathrm{cm^{-3}}$. 
The three doubles $(q_\parallel, \hbar\omega)$ are in each case fixed to the value where 
$P(q_\parallel,\omega)$ is maximal. Since the potential well confining the electrons to 
the film is in our crude model infinitely deep the charge fluctuations are in all cases 
symmetric with respect to the film center. The fluctuation is maximal close to the 
boundaries of the film. Hence, it represents a surface plasmon. Increasing the density 
$n_s^p$ changes mainly the amplitude of the oscillation. The overall structure
remains the same. Hence, we are indeed tracking a particular surface plasmon 
of the film with the density of the surplus electrons coming from the plasma.

At the end of this section let us demonstrate the need for using the nonlocal response theory 
of Mills and coworkers~\cite{Mills75,EM87,SM89}. It is necessary in cases where the screening 
length $\lambda_s=\sqrt{k_BT/4\pi n_b e^2}$ due to the electrons producing the charge fluctuation
the EELS electron couples to is on the same order as the (thermal) de Broglie wave length
$\lambda_{\rm dB}=\sqrt{(2\pi\hbar)^2/3 m_e^*k_BT}$. The density $n_b=n_b^d+n_b^s$ with
$n_b^d$ the bulk electron density due to the background doping and $n_b^p$ the bulk 
density of the electrons coming from the plasma (corresponding to a surface density
$n_s^p= n_b^p d$). As can be seen in Table~\ref{NonLocalTable}, the screening 
length and de Broglie wavelength, given for three different values of $d$ and an 
electron density typical for our setup, are on the same order. The nonlocal theory
is thus required to describe the dielectric response of the electrons in the film. In fact, 
the loss peak we are monitoring is even absent in the local theory which uses a simple Drude 
term added to the film's background dielectric function~\cite{IM82}.

\begin{table}[t]
        \begin{center}
                \begin{tabular}{c|ccc}
                        & \multicolumn{3}{c}{\mbox{\head{\AlTwoOThree}} ($\lambda_{\rm dB}=9.9$\,nm)}\\ \hline
                        d/nm& $n^p_s$[$10^{11}\,$cm$^{-2}]$ & $n_b$[$10^{18}\,$cm$^{-3}]$ & $\lambda_s$[nm]\\ \hline
                        5  &  1 & 1.2  & 3.3 \\
                        10 &  1 & 1.1  & 3.4 \\
                        15 &  1 & 1.07 & 3.5 \\
                \end{tabular}
                \caption{Comparison of the de Broglie wave length $\lambda_{\rm dB}$ and screening length
                         $\lambda_s$ for an \AlTwoOThree\ film with thickness $d$ and charge density
                         $n_b=n_b^d + n_b^p$, where $n_b^d$ is the film's background electron density due
                         to doping, to be taken in all cases as $10^{18}\,\mathrm{cm^{-3}}$, and 
                         $n_b^p=n^p_s/d$ the density of the additional electrons coming from
                         the plasma. As can be seen $\lambda_{\rm dB}$ and $\lambda_s$ are on the same order.
                         Hence, a nonlocal description of the modification of the film's dielectric function
                         due to the charge carriers is required.
                         }
                \label{NonLocalTable}
        \end{center}
\end{table}

\section{Conclusions}
We described an EELS setup for determining the density of electrons accumulated 
by a plasma-facing dielectric solid. It is based on a two-layer structure, 
consisting of a film and a substrate, which can be inserted into the wall of 
a discharge. The film is made out of the material whose plasma-induced charging 
one wants to know while the substrate ensures the stability of the structure and 
the confinement of the charges to the film. It is also the layer to which the probing 
electron beam is applied. The device is geared towards measuring the total charge 
accumulated by the plasma-facing structure. In principle a structure of 
this type could be also used to determine by EELS the profile of the charge 
distribution perpendicular to the plasma-solid interface. It is then however 
necessary to base the theoretical analysis on a selfconsistent kinetic theory of 
the electric double layer at the plasma-solid interface because otherwise the 
width of the space charge cannot be determined. In addition the film has to host the 
whole space charge. How challenging this will be for the EELS sensitivity limit 
future work will show.

The main goal of 
this work was to find an EELS
setup for measuring the total charge residing inside a plasma-facing dielectric film.
For that purpose we made simplifying assumptions and neglected a number of aspects 
which may be of importance for a quantitative analysis of experimental data.
For instance, the plasma in front of the structure is not modelled,
it simply provides surplus charges/electrons for the film. Furthermore, the 
charge confinement is not calculated selfconsistently and the charges inside film 
are assumed to be thermalized. It is also assumed that all the electrons coming 
from the plasma are accumulated in the film's conduction band ignoring surface 
states. These factors will modify the EELS spectrum quantitatively but not 
qualitatively. The principle of our proposal--confining the wall charge to a narrow 
film, stabilizing the film by a substrate, and reading-out the charge density from 
the shift of a loss peak in the from-the-back EELS--is unaffected by them. 

To obtain in the particular scattering geometry on which our proposal is based a
sufficiently strong loss peak, detectable by current EELS instrumentation, it is 
most probably necessary to pre-n-dope the plasma-facing film. For the principle 
of the method the doping is not necessary. The undoped film has loss peaks due 
exclusively to plasma-induced charging but they are rather faint because of their 
multipole character. The pre-doping has however also the nice additional effect of providing 
a reference peak, present also when the plasma is off. Once the plasma is on and the 
film is flooded by electrons from the plasma the peak shifts with the density of the 
additional electrons. From the peak position the density can thus be determined. Our 
results for the \AlTwoOThree/\CaO\ system indicate that the from-the-back geometry 
may indeed work. By a judicious choice of materials--having suitable mechanical 
properties, conduction band edges with appropriate off-sets, and a suite of donors--it 
should be possible to build sub-$100\,\mathrm{nm}$ thick structures for measuring the 
wall charge which are mechanically stable and yet furnish the EELS signal with enough 
strength to be detectable. 

\begin{acknowledgments}
In the initial stages this work was supported by the Deutsche Forschungsgemeinschaft through the Transregional
Collaborative Research Center SFB/TRR24.
\end{acknowledgments}

\bibliography{myref.bib}

\end{document}